\pdfoutput=1

\documentclass[11pt]{article}

\usepackage[final]{acl}

\usepackage{times}
\usepackage{latexsym}
\usepackage{caption}
\usepackage[T1]{fontenc}

\usepackage[utf8]{inputenc}

\usepackage{microtype}

\usepackage{inconsolata}

\usepackage{graphicx}

\usepackage{multirow}
\usepackage{booktabs}
\usepackage{array,booktabs,tabularx} 
\newcolumntype{Y}{>{\centering\arraybackslash}X}
\usepackage{makecell}

\usepackage{xcolor}
\usepackage{pifont}

\definecolor{forestgreen}{RGB}{34,139,34}
\definecolor{steelred}{RGB}{139,30,30}

\newcommand{\cmark}{\textcolor{forestgreen}{\scalebox{1.1}{\ding{51}}}}
\newcommand{\xmark}{\textcolor{steelred}{\scalebox{1.1}{\ding{55}}}}

\newcommand{\yes}{\cmark}
\newcommand{\no}{\xmark}

\usepackage{subfig}  
\usepackage{caption}
\usepackage{adjustbox} 

\usepackage{listings}
\usepackage{xcolor}

\definecolor{jsonkey}{rgb}{0.75, 0.3, 0.1}
\definecolor{jsonstring}{rgb}{0.1, 0.1, 0.75}
\definecolor{jsonnumber}{rgb}{0.0, 0.5, 0.0}

\lstdefinelanguage{json}{
    basicstyle=\ttfamily\footnotesize,
    numbers=left,
    numberstyle=\scriptsize\color{gray},
    stepnumber=1,
    numbersep=6pt,
    showstringspaces=false,
    breaklines=true,
    frame=single,
    backgroundcolor=\color{white},
    literate=
        {true}{{\color{jsonkey}true}}{4}
        {false}{{\color{jsonkey}false}}{5}
        {null}{{\color{jsonkey}null}}{4}
        {:}{{\color{jsonkey}{:}}}{1}
        {,}{{\color{jsonkey}{,}}}{1},
    morestring=[b]",
    morecomment=[l]{//},
    moredelim=[s][\color{jsonstring}]{"}{"},
    moredelim=[l][\color{jsonnumber}]{0},
    moredelim=[l][\color{jsonnumber}]{1},
    moredelim=[l][\color{jsonnumber}]{2},
    moredelim=[l][\color{jsonnumber}]{3},
    moredelim=[l][\color{jsonnumber}]{4},
    moredelim=[l][\color{jsonnumber}]{5},
    moredelim=[l][\color{jsonnumber}]{6},
    moredelim=[l][\color{jsonnumber}]{7},
    moredelim=[l][\color{jsonnumber}]{8},
    moredelim=[l][\color{jsonnumber}]{9}
}

\lstdefinestyle{python}{
    language=Python,
    basicstyle=\ttfamily\footnotesize,
    keywordstyle=\color{blue},
    commentstyle=\color{gray},
    stringstyle=\color{red},
    numbers=left,
    numberstyle=\scriptsize\color{gray},
    stepnumber=1,
    numbersep=6pt,
    frame=single,
    breaklines=true,
    captionpos=b,
    morekeywords={self, with, as, lambda}, 
    showspaces=false,         
    showstringspaces=false    
}

\newcommand{\indoorworld}{\textsc{IndoorWorld }}

\title{\indoorworld: Integrating Physical Task Solving and Social Simulation in A Heterogeneous Multi-Agent Environment}

\author{
  Dekun Wu$^{1}$ \quad
  Frederik Brudy$^{2}$ \quad
  Bang Liu$^{1}$\thanks{\ \ Canada CIFAR AI Chair.} \quad
  Yi Wang$^{2}$ \\
  $^{1}$Université de Montréal \& Mila - Quebec AI Institute \\
  $^{2}$Autodesk Research \\
  \texttt{\{dekun.wu, bang.liu\}@umontreal.ca} \\
  \texttt{frederik.brudy@autodesk.com, ywang485@gmail.com}
}

\begin{document}
\maketitle
\begin{abstract}
Virtual environments are essential to AI agent research. Existing environments for LLM agent research typically focus on either physical task solving or social simulation, with the former oversimplifying agent individuality and social dynamics, and the latter lacking physical grounding of social behaviors. We introduce \indoorworld, a heterogeneous multi-agent environment that tightly integrates physical and social dynamics. 
By introducing novel challenges for LLM-driven agents in orchestrating social dynamics to influence physical environments and anchoring social interactions within world states, \indoorworld opens up possibilities of LLM-based building occupant simulation for architectural design. We demonstrate the potential with a series of experiments within an office setting to examine the impact of multi-agent collaboration, resource competition, and spatial layout on agent behavior. 



\end{abstract}

\section{Introduction}



The emergence of Large Language Model (LLM)-based agents has extended LLMs beyond traditional one-off interactions, equipping them with long-term memory, planning capabilities, and embodied actions~\cite{yao2022react,shinn2024reflexion,wangSurveyLargeLanguage2024}. Among them, multi-agent systems leverage distinct agent roles to achieve greater collective intelligence for problem-solving~\cite{hong2024metagpt,chatdev,tang2024medagents} or to enable more realistic cognitive and psychological modeling in social simulations.

Like in traditional AI research~\cite{maes1995artificial}, virtual environments are essential for LLM-based agents, enabling them to perceive and act. These environments provide external sensory input and introduce a world state, allowing agent actions to be grounded as operators that modify the environment. These environments serve as low-cost testbeds that accelerate LLM-agent development, typically falling into two categories: \textbf{physical task-solving environments} and \textbf{social simulation environments}.

Physical task-solving environments such as VirtualHome~\cite{puig2018virtualhome} and ALFWorld~\cite{shridhar2020alfworld} enable agents to interact with external objects to manipulate the world state towards specific objectives.
However, these environments often assume identical action spaces, homogeneous agent abilities, neglecting individual differences among agents, and the impact of social dynamics on task solving.


On the other hand, social simulation environments, such as Smallville~\cite{Park2023GenerativeAgents}, enable social interactions between agents, such as relationship building and information sharing. While these systems effectively simulate human-like social behaviors driven by individual personalities and roles, they often employ an oversimplified model of the physical world. This leads to a lack of groundings for agents' actions in the change of world state, resulting in social interactions that remain merely  ``plausible'' without any correspondence to an external physical reality. For example, an agent may refer to non-existent physical objects in a dialog.




The gaps lead to missed opportunities to use AI agents for applications that require tight integration of physical task-solving and social simulation. One such application is building occupant simulation for architectural design \cite{yan2015occupant,feng2015simulation}, where occupant behaviors are driven by both dynamic physical and social factors. To respond to the missed opportunities, we present \indoorworld, a heterogeneous multi-agent environment that tightly integrates physical and social dynamics in an indoor space setting, 
introducing novel challenges for LLM-driven agents in orchestrating social dynamics to influence physical environments and anchoring social interactions within physical world states. As a multi-agent system, \indoorworld allows for fully decentralized agent control and collective task assignment, facilitating self-regulated labor division, task prioritization and coordination. \indoorworld provides a scalable and expressive testbed for advancing research on multi-agent LLM systems.


Our key contributions are as follows:
\begin{itemize}
\item \textbf{Heterogeneous Agent Modeling:} We introduce a multi-level approach in which agents vary in roles, actions, capabilities, and knowledge, yielding individual differences that professionals judged to be more realistic.

\item \textbf{Integrated Physical and Social Dynamics:} Our environment seamlessly combines physical object manipulation with social behaviors, posing novel challenges and setting the stage for developing LLM-based multi-agent systems for both task solving and social simulation.

\item \textbf{Promising Tools for Architectural Design:} Experiments on multi-agent collaboration, resource competition, and layout effects demonstrate that our platform can aid spatial optimisation and resource allocation, making it a promising tool for architectural design work.

\end{itemize}

\begin{figure*}[h]  
    \centering
\includegraphics[width=1\textwidth]{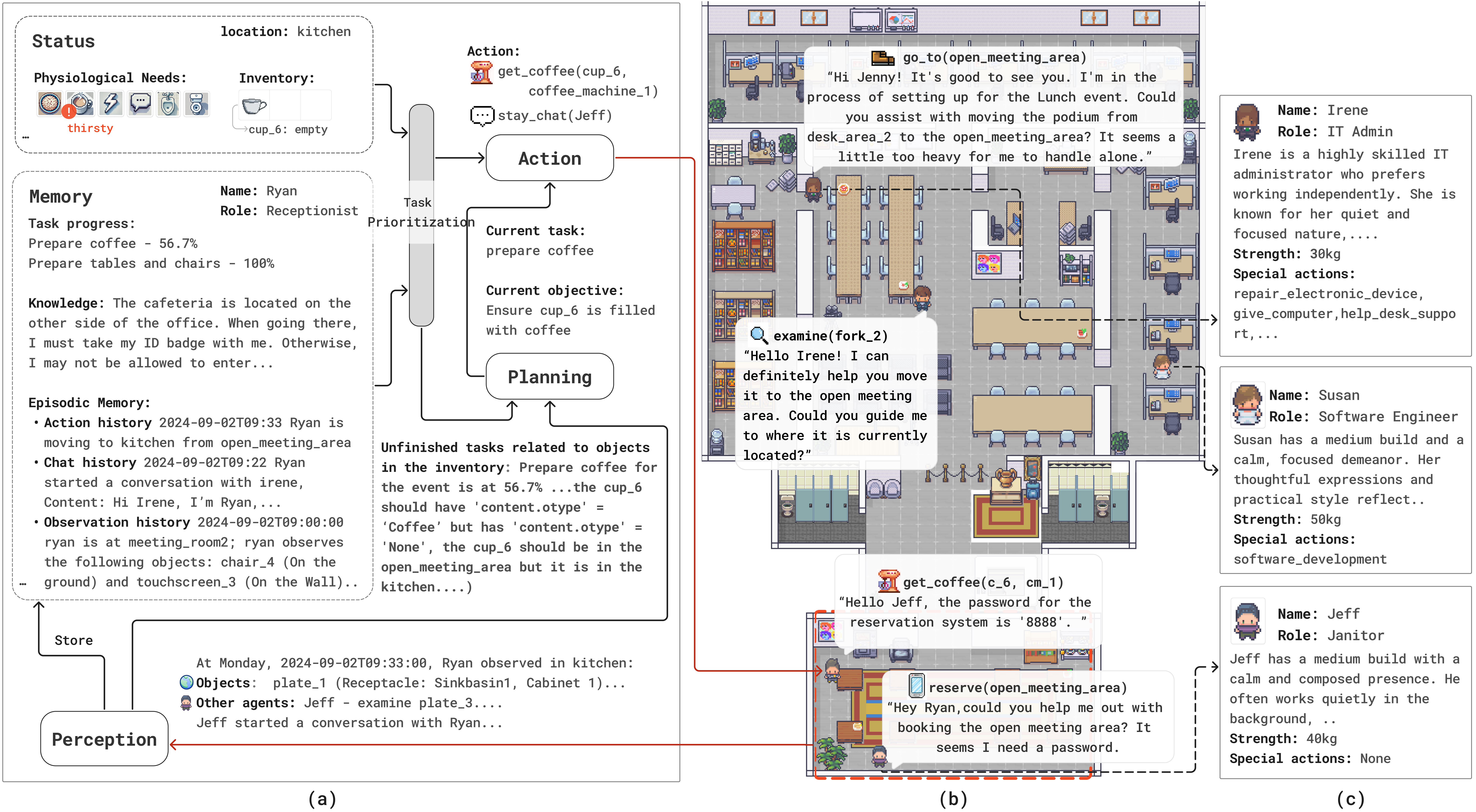}
    \caption{\indoorworld system. (a) Agent architecture; (b) Example agent behaviors; (c) Example of heterogeneous agent profiles}
    \label{fig:agent}
\end{figure*}

\section{Related Works}

\paragraph{Task Solving Environments for LLM-based Agents} evaluate agents' ability to solve various types of tasks, such as web-based tasks~\cite{cai2024personalwab,chae2024webagentsworldmodels}, GUI tasks~\cite{nguyen2024gui,wang2024oscar}, coding tasks~\cite{hong2024metagpt,huang2024agentcoder,chatdev} and household tasks~\cite{ALFWorld20,ALFRED20,zhang2024building}. In household task-solving environments, agents must explore their surroundings, sense and interpret object states, plan and execute actions. The advantage of such environments lies in their support for a diverse range of physical objects and extensive agent-object interactions.
For example, ALFWorld~\cite{ALFWorld20} enables agents to interact with objects such as mugs, books, and lamps. Multi-agent platforms like TDW-MAT and VirtualHome~\cite{zhang2024building,puig2018virtualhome} supports interactions with objects like pens, beds, and apples. Similarly, MineLand~\cite{yu2024mineland} and AdaSociety~\cite{huang2025adasociety}, designed for wilderness survival, feature various tools and food items. However, a common limitation across these environments is the lack of explicit modeling for agent heterogeneity. ALFWorld is a single-agent environment, while TDW-MAT and VirtualHome~\cite{puig2018virtualhome} features homogeneous agents with identical capabilities. Although inventory variations in AdaSociety and MineLand introduce some level of heterogeneity, the agents remain fundamentally homogeneous, as they share the same action space.

\paragraph{Social Simulation Environments for LLM-based Agents} enable agent-agent interactions, elevating the importance of social dynamics. These environments often model differences in personality, profession, and other traits among agents~\cite{Park2023GenerativeAgents,wu-etal-2024-deciphering,xu2023exploring,guan2025richelieu,li2024econagent}, as well as incorporate human needs modeling~\cite{wang-etal-2023-humanoid,wang2024simulating}. However, a major limitation of these environments is the oversimplified modeling of the physical world, including interaction with physical objects.
For instance, an agent may perform the action of eating, without any explicit modeling of food items in the environment. 
This prevents these simulation environments to be applied in settings where the physical environment has impacts on agent behaviors, such as resource allocation and layout study.

\paragraph{AI in Architectural Design} has been widely adopted, mainly to generate static 2D and 3D artefacts, such as floorplans, interiors, and furniture layouts, but these visually oriented methods provide little insight into the dynamic occupant activities~\cite{li2025generative,infinigen2024indoors,leng-etal-2023-tell2design}. Moreover, traditional occupant-simulation tools rely on rule-based or state-based behavior transitions~\cite{schaumann2017simulating,lee2021designing,liu2024simulating}, offering far less flexibility and realism than LLMs in responding to changing environments. Our work addresses this gap by employing LLMs to simulate multi-agent interactions within indoor spaces, yielding insights into occupant behaviors that are crucial for effective spatial planning and resource management.


\vspace{2mm}

\section{\indoorworld Environment}






Our environment \indoorworld seamlessly integrates task-solving and social simulation in an indoor space setting, offering a versatile testbed to study the interplay between social and physical agent behavior. \indoorworld enables agents with rich individual differences to collaborate on tasks with complex hierarchy, involving self-regulated labor division, prioritization and coordination, while satisfying physiological needs through environmental interactions. By incorporating fine-grained modeling of agent internal structure and decision-making process, \indoorworld more accurately resembles real-world human behaviors. Tables~\ref{tab:environment_comparison}, \ref{tab:heterogeneity_comparison}, and \ref{tab:number_of_actions} summarize the key differences between \indoorworld and existing platforms, highlighting its potential to advance research in LLM-based multi-agent systems.

\subsection{Environment Architecture}



\paragraph{Framework Overview and Core Components}

Virtual environments that support planning and task solving often require managing the world state~\cite{srivastava2021behavior,ALFWorld20}. Similarly, our \indoorworld adopts an object-oriented approach to define state transition systems through three key components: 1) \textbf{Agents}, 2) \textbf{Objects}, and 3) \textbf{Locations}.

Each agent and object is associated with state variables, such as a numerical value for an agent’s hunger or a Boolean indicating whether a computer is broken. A dedicated variable tracks each entity’s current location. The overall \textbf{world state} is the joint valuation of all these variables. Object affordances are defined by the set of actions that can be performed on them, and these actions are further constrained by the agent’s role (e.g., only an IT admin can repair computers). When actions are executed, the world state updates accordingly. 
Objects can be marked as {\em receptacles} to store other objects. Locations can be interconnected to allow agent and object movement.


Agent interactions are through conversations. \indoorworld supports 4 actions related to conversation with other agents: 1) ${\tt initiating\_chat}$; 2) ${\tt stay\_chat}$; 3) ${\tt end\_chat}$, and 4) ${\tt join\_chat}$. We let the LLMs to generate free-form dialog content. Note that we allow any number of agents to be in a conversation, as long as they are at the same location. A dedicated agent state variable indicates which conversation session the agent is currently involved (if any). Action 2) and 3) only become admissible actions when the agent is in a conversation, and 4) becomes admissible when there is an ongoing conversation session at the agent's current location. Each agent can only be at one conversation session at a time. Conversations can be used to share information (including task progress), discuss labor division, and coordinating actions. They affect the agent's subsequent actions by updating the agent's internal state. Note that in task-solving scenarios, conversations are \textbf{utility-driven}---the dialog content need to serve task-solving. This aspect distinguishes our work from most existing environments featuring agent conversations.

\begingroup
\setlength{\tabcolsep}{2pt}  
\renewcommand{\arraystretch}{0.9}  
\begin{table}  
\centering
{\footnotesize  
\begin{tabular}{l|cccccccc}
\hline
 & MA & AT & OI & TE & LS & FH & HN & RF \\
\hline
\makecell[l]{ALFworld\\\cite{ALFWorld20}}  
   & \no  & --  & \yes  & \yes  & \no  & \no  & \no  & \yes  \\
\makecell[l]{Virtual-Home\\\cite{puig2018virtualhome}}  
   & \yes & Ho.  & \yes  & \yes  & \yes  & \no  & \no  & \yes  \\
\makecell[l]{TDW-MAT\\\cite{zhang2024building}}  
   & \yes & Ho.  & \yes  & \yes  & \no   & \no  & \no  & \yes  \\
\makecell[l]{C-WAH\\\cite{zhang2024building}}  
   & \yes & Ho.  & \yes  & \yes  & \no   & \no  & \no  & \yes  \\
\makecell[l]{MineLand\\\cite{yu2024mineland}}  
   & \yes & Ho.  & \yes  & \yes  & \yes  & \no  & \yes  & \no  \\
\makecell[l]{AdaSociety\\\cite{huang2025adasociety}}  
   & \yes & Ho.  & \yes  & \yes  & \yes  & \no  & \yes  & \no  \\
\makecell[l]{Smallville\\\cite{Park2023GenerativeAgents}}  
   & \yes & He.  & Lmtd. & \no   & \yes  & \no  & \no  & \yes  \\
\makecell[l]{Humanoid Agents\\\cite{wang-etal-2023-humanoid}}  
   & \yes & He.  & Lmtd. & \no   & \yes  & \no  & \yes  & \yes  \\
   \hline
\makecell[l]{\textbf{Ours}}  
   & \yes & \textbf{He.} & \yes  & \yes  & \yes  & \yes  & \yes  & \textbf{\yes}  \\
\hline
\end{tabular}}
\caption{Comparison of environments. MA: Multi-Agents, AT: Agent Type, OI: Object Interaction, TE: Task Eval, LS: Life Simulation, FH: Fine-Grained Heterogeneity, HN: Human Needs, RF: Real-world Fit. "Ho." = Homogeneous, "He." = Heterogeneous, "Lmtd." = Limited.}
\label{tab:environment_comparison}
\vspace{-5mm}
\end{table}
\endgroup


\paragraph{Sessions and Scenarios}
\indoorworld supports both \textbf{task solving} and \textbf{simulation} sessions. In a task-solving session, a set of tasks are assigned as the shared objective for all agents. The agents need to collectively decide on task orders and assignments, as well as planning for specific action sequences to complete each task. In simulation sessions, no explicit objective is defined.

Both session types start with an initial world state defined by a \textbf{scenario}, a JSON configuration file specifying agents, objects, locations, inter-location connections, and receptacle assignments. (See Appendix~\ref{sec:appendix_b} for an example JSON file.)


To facilitate experimentation, \indoorworld comes with 25 predefined object types (including 7 receptacle types) and 4 predefined agent roles, resulting in a total of 38 action types. The current set of objects and agent roles cover typical activities in an office environment, such as booking meeting rooms, moving desks, cleaning utensils, repairing computers, etc. As shown in Table~\ref{tab:number_of_actions}, our environment offers a broader action space compared to many existing text-based and 2D/3D platforms. 

\paragraph{Customization and Expansion}


Although \indoorworld currently feature a limited number of object types and agent roles, our object-oriented approach allows easy customization of object type and agent roles. Introducing new object types and agent roles involves defining Python functions specifying new interactions, including preconditions and effects, which can be easily achieved by utilizing existing class hierarchy.
We provide example code in Appendix~\ref{sec:appendix_c} to illustrate how to introduce new object types, agent roles and interaction type.

\subsection{Agent Architecture}
\label{subsec:agent_arch}

Figure~\ref{fig:agent} (a) illustrates the operational framework of our agents, highlighting their interactions with both the environment and other agents. Our architecture integrates cognitive modules inspired by recent research on LLM-based agents \cite{yao2022react,zhang2024building} and consists of five core modules: perception, memory, planning, action, and task prioritization.

\paragraph{Perception:} This module processes symbolic observations from the environment, such as nearby objects, receptacles, and other agents' activities, and updates the agent’s internal state accordingly.

\paragraph{Memory:} The memory module stores agent-specific information, long-term knowledge, and interaction history. Inspired by the COELA model \cite{zhang2024building}, it maintains a semantic map and tracks task progress, while also recording episodic events (e.g., past actions and conversations) and retaining pre-existing knowledge. It further monitors internal states like physiological needs and inventory status.

\paragraph{Planning:} Using current observations and stored memories, the planning module determines the agent’s current objective and task to address both task-specific and internal needs.

\paragraph{Action:} Informed by the ReAct framework \cite{yao2022react}, the action module integrates information from perception, memory, and planning to select and execute the next action. This process involves first reasoning about the current situation and evaluating available options before deciding the next move\footnote{The reasoning part is omitted in Figure~\ref{fig:agent} due to space limitations.}.

\paragraph{Task Prioritization:} Our preliminary experiments revealed that LLM-based agents, particularly those using open-source models, struggle to maintain focus in multi-task scenarios, frequently switching tasks without completing them (See Sec.~\ref{subsec:results}). To address this, we proposed a task prioritization module that encourages agents to concentrate on ongoing tasks. The module monitors the objects an agent holds and their relevance to the current task, reminding the agent of incomplete objects. For example, as shown in Figure~\ref{fig:agent}, when agent Ryan is working on preparing coffee and is carrying an empty \texttt{cup\_6}, the module highlights its incomplete status, such as the absence of coffee or its incorrect placement. When no active task is detected, the module reminds the agent about all unfinished tasks, guiding the agent to select an objective aligned with its role and skills. This approach promotes concentration on ongoing tasks and minimizes inefficient task switching. Note that the module does not introduce extra information but selectively reiterates relevant parts of the agent's memory, such as task progress, to reinforce task awareness.

\paragraph{Modeling Agent Heterogeneity}
\indoorworld addresses the limited diversity found in existing multi‑agent benchmarks (Table~\ref{tab:environment_comparison} and ~\ref{tab:heterogeneity_comparison}) by assigning every agent a \emph{multi‑level profile}.  
Each profile combines a \textit{role} that determines the agent’s unique action space (e.g.\ only IT administrators can \texttt{repair} devices) with additional attributes such as \textit{personality}, \textit{strength}, \textit{skill}, and \textit{knowledge}.  
This layered design (illustrated in Fig.~\ref{fig:agent}c) supports emergent division of labour and coordinated behaviors while greatly increasing realism.

\noindent We model agent heterogeneity at \textbf{four levels}: (i) \emph{profile level}, capturing differences in personas and role configurations; (ii) \emph{action space}, where roles have distinct actions; 
(e.g., only IT admins can \texttt{repair} devices);
(iii) \emph{capability}, meaning agents may perform the same action with different efficiency or outcomes, such as janitors cleaning more quickly or strong agents being able to move heavy objects; and (iv) \emph{knowledge}, whereby agents hold different internal information (e.g., only receptionists know how to book a meeting room).


\noindent

To validate this design, 20 practicing architects rated whether heterogeneity at each level increases realism and whether it is important
for understanding real space use (details in Appendix~\ref{sec:appendix_d}).
The consistently high realism scores (55--70\,\%) and substantial importance ratings
(55--95\,\%) demonstrate that multi‑level heterogeneity is both credible and
valuable for architectural analysis, thereby supporting the soundness of our design choices.

\paragraph{Modeling Human Needs} In \indoorworld, 
agents are associated with physiological and social needs, such as hunger, thirst, and social interaction, that resemble actual building occupants (Figure \ref{fig:agent}).
They are tracked with numerical state values that gradually decline over time, prompting agents to perform restorative actions like eating, drinking, socializing, or using the restroom.
This explicit modeling fosters more natural and realistic behaviors as agents balance personal upkeep with assigned tasks.

\section{Experiments}

\subsection{A Collaborative Task Solving Benchmark}

To demonstrate how \indoorworld supports task solving with social interactions, we design a benchmark for collaborative task solving in an office setting. 
The benchmark consists of an overarching task: preparing for a company event, composed of five major subtasks, each involving common office activities such as transporting items, cleaning utensils, repairing equipment, booking meeting spaces, and preparing food and beverages. 
These subtasks can be broken down into smaller steps that require coordination among agents. Unlike many existing environments where agents execute a single task in isolation \cite{ALFWorld20}, in our setting, all agents simultaneously receive the complete task structure,
but must autonomously determine how to divide responsibilities, prioritize actions, and coordinate efforts in a decentralized manner. This introduces additional challenges related to task prioritization, division of labor, communication, and coordination strategies. Each scenario runs for one hour of simulation time, during which agents must collaboratively complete as much of the task as possible. The scenarios feature 9 locations, 67 objects across 16 types (including 15 receptacles of 7 types), and 6 agents: four janitors, one IT administrator, and one receptionist. Details of each subtask and its decomposition can be found in Appendix~\ref{sec:appendix_a}.

\subsection{Social Simulation Experiments}


Like Smallville~\cite{Park2023GenerativeAgents}, our environment supports autonomous social simulation where agents generate their own objectives based on physiological/social needs, personality, and roles, driving interactions with both the environment and other agents.


As \indoorworld simulates both social and physical interactions, one potential real-world application is to evaluate how well a physical environment can facilitate the activities of its occupants. We demonstrate this application with two experiments, focusing on resource management and spatial layout design, respectively. 

\paragraph{Resource Management} experiment showcases how we can use \indoorworld to evaluate resource allocation in an environment. The scenario involves agents residing in a space with limited resource. We examine how agents compete for and utilize resources under different conditions.

We ran 3 sets of simulations, with 2, 4, and 8 thirsty agents. Each set consist of three settings involving agents all preferring water, all preferring coffee, and no preference, and for each setting we ran simulation with different availability of beverages (one water dispenser/coffee machine vs. two water dispensers/coffee machines).

At the beginning of the simulation, agents experience thirst and autonomously decide whether to drink water or coffee based on their physiological needs and personal preferences. We analyze their resource selection behavior and measure the time required for all agents to fulfill their hydration needs~\footnote{Unless otherwise specified, all time measurements in this paper refer to in-simulation time.}. This allows us to assess how different resource configurations influence competition and overall efficiency in meeting agent needs.

\begin{figure}[htbp]  
    \centering
        \caption{Spatial layout of design 1 and design 2}
    \includegraphics[width=0.55\textwidth]{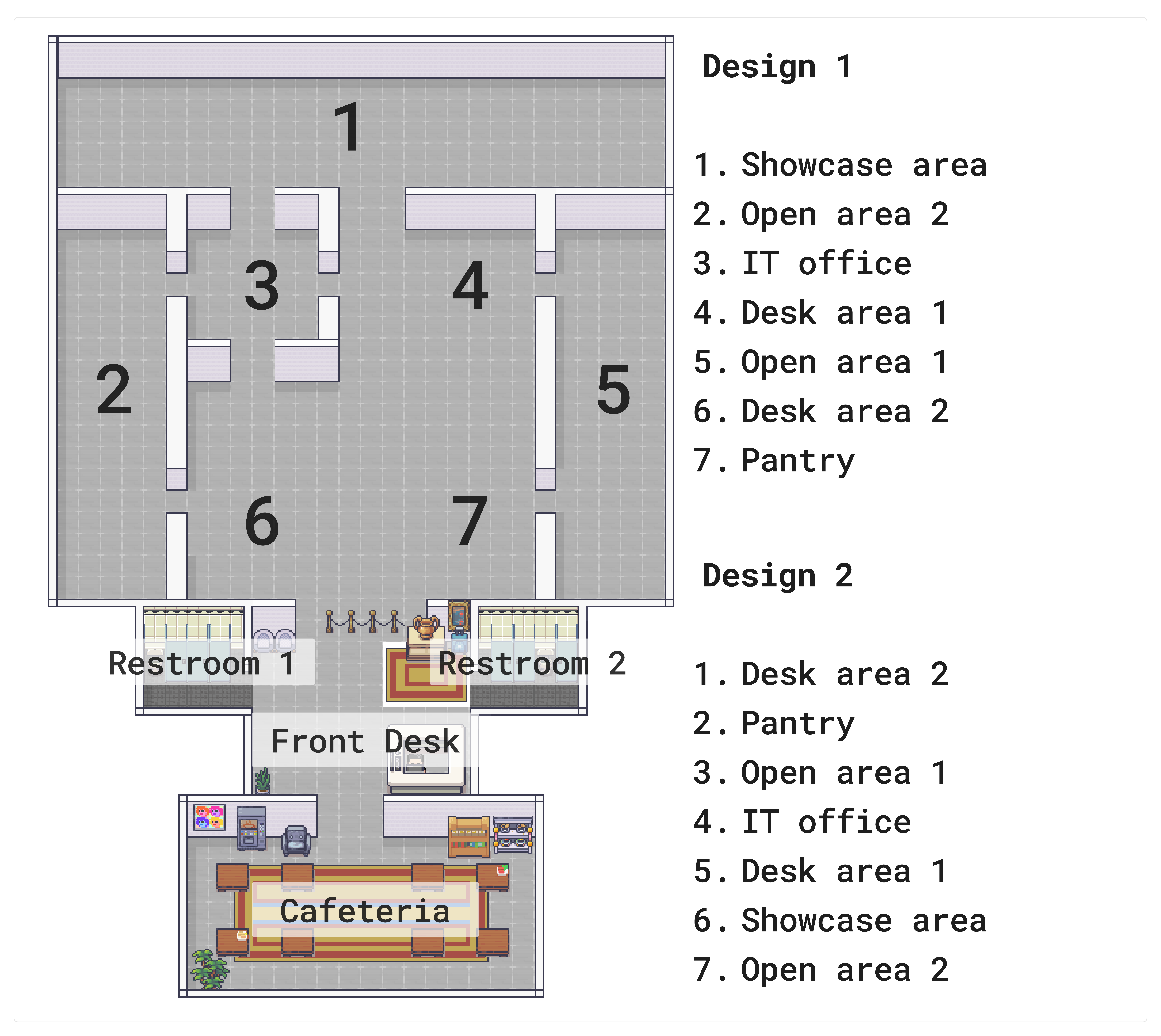}

    \label{fig:spatial_layout}
     \vspace{-5mm}
\end{figure}

\paragraph{Spatial Layout} experiment showcases how we can use \indoorworld to evaluate building layout designs. We ran simulations in two scenarios involving two office layout designs (see Figure \ref{fig:spatial_layout}), each lasting for an 8-hour simulation period. Both Design 1 and Design 2 contain the same 11 functional areas to host five agents: one IT administrator, one janitor, two software engineers, and one receptionist. Each agent has a designated workspace, such as the front desk for the receptionist and the IT office for the IT administrator.

Furthermore, the pantry contains a coffee machine, a water dispenser, food, drinks, cups, and utensils, but with limited resources. The cafeteria, in contrast, is assumed to have no resource constraints, allowing agents to replenish food, drinks, and energy freely. However, activities in the cafeteria takes more time compared to the pantry. Additionally, the environment includes male and female restrooms. 

In both Design 1 and Design 2, all areas share the same configurations, including resource availability and the activities they support. The only difference between the two designs lies in the relative positioning of these areas.


Since this experiment does not involve specific task execution, the task prioritization and task progress modules have been removed.

\begin{table*}[h!]
\centering
\footnotesize
\caption{Results of collaborative task solving. T1-T5 represent the five designed tasks. }
\begin{tabular}{@{}lcccccc@{}}
\toprule
 & \textbf{T1} & \textbf{T2} & \textbf{T3} & \textbf{T4} & \textbf{T5} & \textbf{AVG} \\
 & (IS/AS) & (IS/AS) & (IS/AS) & (IS/AS) & (IS/AS) & (IS/AS) \\
\midrule
Random Agents 
& 8.3/38.9 
& 0.0/30.6
& 0.0/30.0
& 0.0/0.0
& 0.0/11.1
& 1.2/23.1 \\
\midrule
\multicolumn{7}{c}{\textbf{Llama 3.3 70B Instruct}} \\ 
\midrule
Full Model w/o S\&T and TP 
 & 83.3/88.9 & 38.9/61.1 & 0.0/53.3 & 100.0/100.0 & 0.0/27.8 & 31.0/51.8 \\
Full Model w/o TP  
 & 91.7/94.4 & 36.1/59.3 & 0.0/53.3 & 100.0/100.0 & 0.0/41.1 & 31.0/56.1 \\
Full Model
 & \textbf{100.0}/\textbf{100.0} & \textbf{58.3}/\textbf{72.2} & \textbf{33.3}/\textbf{73.3} & 100.0/100.0 & \textbf{33.3}/\textbf{51.1} & \textbf{55.2}/\textbf{67.8} \\
\midrule
\multicolumn{7}{c}{\textbf{Gemma 3 27B}} \\
\midrule
Full Model w/o S\&T and TP   & 75.0/83.3 & 11.1/45.4 & 33.3/73.3 & 100.0/100.0 & 12.5/40.0 & 26.4/51.4 \\
Full Model w/o TP 
& 50.0/66.7 & 50.0/70.4 & 0.0/60.0 & 33.3/33.3 & 0.0/30.0 & 28.7/53.3 \\
Full Model 
& \textbf{83.3}/\textbf{88.9} & \textbf{72.2}/\textbf{81.5} & 25.0/70.0& 66.7/66.7 & \textbf{45.8}/\textbf{65.6}& \textbf{59.8}/\textbf{74.5} \\
\midrule
\multicolumn{7}{c}{\textbf{GPT-4o}} \\
\midrule
Full Model w/o S\&T and TP 
& 75.0/83.3 & 52.8/77.8 & 0.0/50.0 & 66.7/66.7 & 33.3/58.9 & 43.7/67.8 \\
Full Model w/o TP 
& 91.7/94.4 & 77.8/85.2 & 66.7/86.7 & 100.0/100.0 & 62.5/76.7 & 74.7/83.5 \\
Full Model 
& \textbf{100.0}/\textbf{100.0} & 61.1/74.1 & \textbf{100.0}/\textbf{100.0} & 100.0/100.0 & \textbf{83.3}/\textbf{87.8} & \textbf{79.3}/\textbf{84.7} \\
\bottomrule
\end{tabular}
\vspace{-3mm}
\label{table:collaborative_task_solving}
\end{table*}

\begin{table}[h!]
\centering
\footnotesize
\caption{Facility Resource Stress Testing. The second row (1/1, 2/2) indicates the number of water dispensers and coffee machines. The left column (2, 4, 8) represents the number of agents. Each cell (X/Y/Z) shows the number of agents who drank water (X) and coffee (Y), and the total time (Z) for all agents to hydrate.}
\begin{tabularx}{0.48\textwidth}{@{}lYYYYYY@{}}
\toprule
\multirow{2}{*}{} 
  & \multicolumn{2}{c}{\textbf{All Like Water}} 
  & \multicolumn{2}{c}{\textbf{All Like Coffee}}
  & \multicolumn{2}{c}{\textbf{No Preference}} \\ 
\cmidrule(lr){2-3} \cmidrule(lr){4-5} \cmidrule(lr){6-7}
& 1/1 & 2/2 & 1/1 & 2/2 & 1/1 & 2/2 \\ 
\midrule
2 & 2/0/4 & 2/0/3 & 0/2/4 & 0/2/3 & 2/0/4 & 2/0/3 \\ 
4 & 4/0/6 & 4/0/4 & 0/4/6 & \textbf{1}/\textbf{3}/\textbf{4} & 4/0/6  & 4/0/5 \\ 
8 & 8/0/10 & 8/0/7 &0/8/10 &0/8/7 & 8/0/12 & 8/0/7 \\ 
\bottomrule
\end{tabularx}
\label{tab:resource_stress}
\vspace{-3mm}
\end{table}

\subsection{Results and Analysis}
\label{subsec:results}
In the collaborative task solving experiment, we evaluated three different LLMs, including the open-source Llama 3.3 70B Instruct, Gemma 3 27B and the proprietary GPT-4o-08-06. For the social simulation experiments, the resource management experiment used GPT-4o-08-06, while the spatial layout experiment used GPT-4o-mini-0718. The temperature for all experiments was set to 0.6.

\paragraph{Collaborative Task Solving} Quantitative results are shown in Table~\ref{table:collaborative_task_solving}. We report {\em instance-level} (IS) task completion rate (percentages of objects that are at the desired state) and {\em attribute-level} (AS) task completion rate (percentages of object state variables that are at the desired value). Removing task prioritization (TP) led to performance degradation across all models, with the most significant drop observed in Llama 3.3 and Gemma 3. This suggests that these models struggle with maintaining focus in multi-task scenarios and benefit substantially from explicit prioritization to reduce inefficient task switching. Further removing the semantic map and task progress tracking (S\&T) led to additional performance degradation. GPT-4o exhibited a substantial drop in performance, indicating that it can effectively leverage the additional structured information for task planning and execution. In contrast, the removal of S\&T had a more limited impact on Llama 3.3 and Gemma 3, likely due to their weaker information utilization capabilities. These models may struggle to extract key insights from the complex semantic map and task progress data, making it difficult for them to identify and prioritize the most relevant information. This observation suggests that while structured task representations are beneficial, their effectiveness is contingent on the model’s ability to process and utilize the provided information efficiently.




In the experiments, we observe various collaborative behaviors among the agents.
The collaboration and labor division reflects individual differences across agents. For example, Irene (IT Admin) asked Jenny (Janitor) for help moving a podium beyond her strength. (Figure \ref{fig:agent} (b)). Collaboration does not only involve physical interaction, but also information sharing via communication. In a different session, Jeff is trying to reserve a meeting room, for which a password is needed. Initially, only the receptionists have knowledge about the password. Jeff first attempted a random password (hallucinated by LLMs). After failing, he asked the receptionist Ryan for the password, and successfully booked the meeting room (Figure \ref{fig:agent} (b)). 


We observe that without task prioritization module, agents tend to switch back and forth between different tasks. For example, Jake is in the kitchen with an uncleaned ${\tt fork\_3}$ in his inventory. As a janitor, Jake should clean them as part of Task 2. However, he suddenly remembers that ${\tt chair\_4}$ needs to be moved to the open area 1 for Task 1, so he dropped ${\tt fork\_3}$ to pick up ${\tt chair\_4}$. But afterwards he recalls the former task again and picks ${\tt fork\_3}$ again. The behavior repeated and resulted in Jake not achieving any goals.

Admittedly, even for models with a high completion rate, tasks may not always be completed in the most efficient way. For example, a receptionist may choose to clean the dishes themselves instead of asking for help from the janitor, who can clean dishes much faster. Lack of global coordination also sometimes leads to duplicated task completion by agents in different locations. For instance, Irene was unaware that her colleagues had already prepared enough tea in the open area 1, so she continued preparing tea in the kitchen and attempted to carry it over. Optimizing collective task completion efficiency remains a challenge.

\paragraph{Resource Management Results} are shown in Table~\ref{tab:resource_stress}. We observed that increasing the number of agents led to longer hydration times due to increased competition and queuing for resources. However, when the number of available water dispensers and coffee machines was increased, agents were able to hydrate more quickly, demonstrating that more resources can alleviate competition and improve efficiency.

We also found that agent preferences significantly influenced their resource selection. When all agents preferred water, they consistently chose the water dispenser, even when a coffee machine was available. Similarly, when all agents preferred coffee, they almost exclusively used the coffee machine, ignoring the water dispenser. In the no-preference scenario, all agents opted for the water dispenser. This may suggest that in the absence of a strong preference, LLM-based agents default to an prior knowledge that biased towards water as the primary hydration method. This may also explain why, even when all agents preferred coffee, a small number of them still chose the water dispenser.

As can be seen, simulation in \indoorworld can effectively reflect the impact of different resource allocation strategies, making it potentially a useful tool to aid real-world decision making on resource allocation.

\begin{figure}[htbp]
    \centering
    \includegraphics[width=1\linewidth]{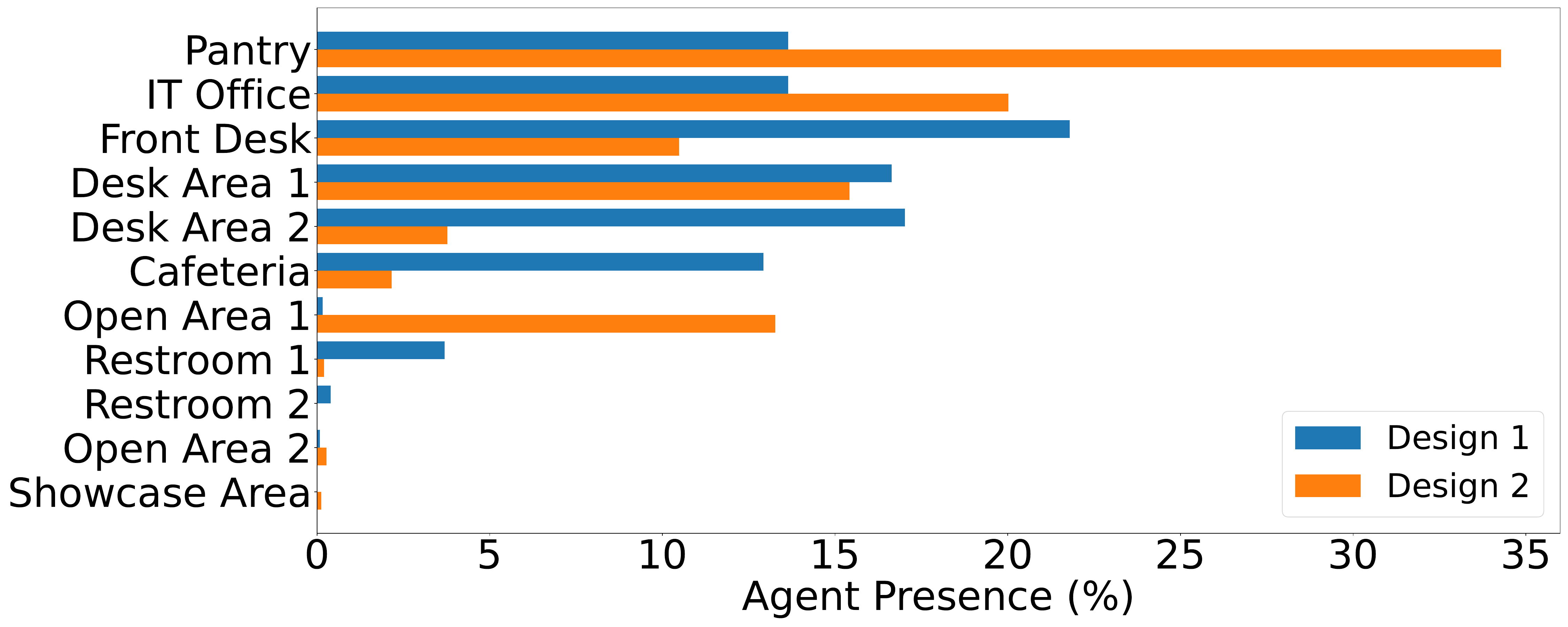}
    \vspace{-5.5mm}
    \caption{Agent Presence Across Locations.}
    \label{fig:agent_distribution}
\end{figure}

\vspace{-5.5mm}
\begin{figure}[htbp]
    \centering
    \includegraphics[width=1\linewidth]{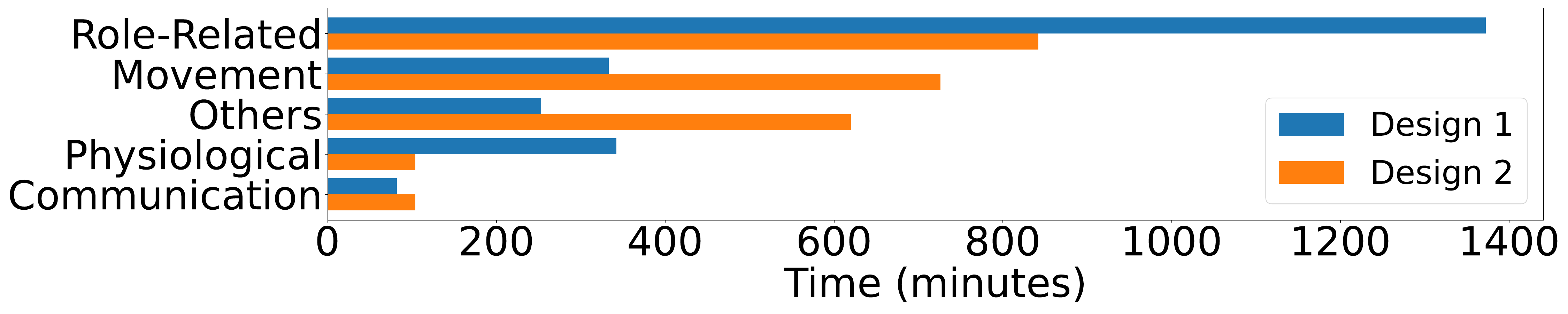}
    \vspace{-3.5mm}
    \caption{Activity Time Distribution.}
    \label{fig:activity_distribution}
\end{figure}

\vspace{-5.5mm}

\begin{figure}[htbp]
    \centering
    \includegraphics[width=1.0\linewidth]{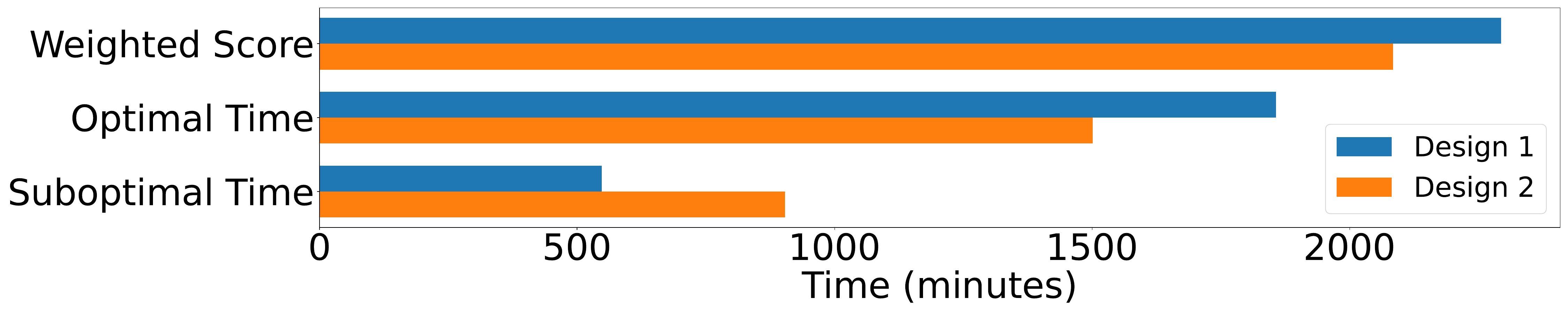}
    \vspace{-8mm}
    \caption{Agent Well-being Metrics.}
    \label{fig:agent_wellbeing}
\end{figure}

\vspace{-1mm}
\begin{figure}[htbp]
    \centering
    \includegraphics[width=0.98\linewidth]{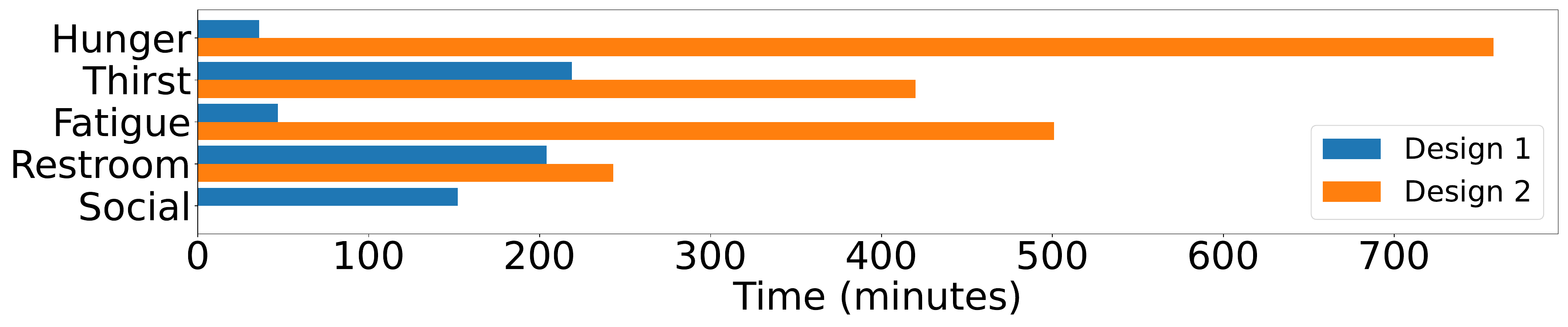}
    \vspace{-3.5mm}
    \caption{Suboptimal State Distribution.}
    \label{fig:suboptimal_conditions}
    \vspace{-3mm}
\end{figure}

\paragraph{Spatial Layout Results} illustrates the impact of different spatial layouts on agent behavior, focusing on location usage, activity time allocation, and overall well-being. The results indicate that spatial design significantly influences resource accessibility, social interactions, and agent efficiency and well-being.

As shown in Figure~\ref{fig:agent_distribution}, Design 1 maintained a more balanced distribution of agents across locations, while Design 2 saw a higher concentration in resource-rich areas, such as the pantry, leading to reduced workspace utilization, particularly for Desk Area 2, which is positioned in the farthest corner. This suggests that inefficient resource distribution may cause agents to move away from their designated workspaces, opting instead to work or engage in activities elsewhere.

Figure~\ref{fig:activity_distribution} illustrates that agents in Design 2 spent more time moving and less time on role-related work due to longer distances from their workspace to resource-rich areas. Social interaction time was also higher, while time spent addressing physiological needs was lower. According to the simulation log, in Design 2, agents spent a significant portion of their time in the pantry engaging in conversations. As a result, despite their prolonged stay in the pantry, the time spent on physiological needs-related activities remained relatively low.

Figure~\ref{fig:agent_wellbeing} and Figure~\ref{fig:suboptimal_conditions} further show that agents in Design 2 spent less time in an optimal state and more time experiencing unmet needs, particularly hunger and fatigue. This can be attributed to the limited resources available in the pantry, such as only two pieces of bread, two apples and a few clean cups, which were quickly consumed.

\paragraph{Implications for Architectural Design}
To evaluate how \indoorworld can support professional practice, we surveyed 9 architects and summarized their feedback in Table~\ref{tab:design_relevance_usefulness}. The results indicate that our features (human need modelling, resource management, and spatial layout experiments) align closely with the considerations architects make during design, and the spatial layout experiments were rated as especially helpful for understanding occupant behavior. Results of the survey and discussions are provided in Appendix~\ref{appendix:survey_2}.

\section{Conclusion}

This study introduces a \indoorworld, a multi-agent simulation environment that integrates fine-grained heterogeneous agent modeling with physical interactions. \indoorworld enables agents with varied abilities to coordinate roles, satisfy physiological needs, and interact with a rich array of objects within realistic settings.

We evaluate three LLMs in collaborative task solving using a benchmark of common office tasks that involve both social and physical interactions, highlighting the effectiveness of task prioritization. Our simulation experiments on resource management and spatial layouts demonstrate the potential real-world applications of \indoorworld in architectural design.

\newpage
\section*{Limitations}

\paragraph{Impact of Experimental Variability on Conclusions}
Due to the inherent randomness of LLMs, LLM-based agents may make different decisions even when facing identical scenarios. Furthermore, since our environment involves multiple agents making sequential decisions across multiple rounds, and their choices influence one another, the results may vary even when using the same experimental settings and LLM model. In our experiments, we observed that fluctuations in results were often associated with the following phenomena:  
(a) Agents engaging in prolonged conversations without progressing on task completion.  
(b) Agents performing incorrect actions after making substantial progress, leading to task failure.  
For instance, in one run of the GPT-4o Full model without S\&T and TP (see Table~\ref{table:collaborative_task_solving}), an agent unexpectedly moved a table, already placed in the open area 1, back to the kitchen. As a result, both the table and the clean utensils on it were relocated to an incorrect position, significantly lowering the overall task success rate.  

To mitigate the impact of experimental variability on the reliability of conclusions, we conducted three independent runs for each task-solving experiment. Additionally, we commit to publicly releasing our code after acceptance to enable further reproducibility and facilitate research on LLM-based agents’ behavior, reasoning capabilities, and collaborative problem-solving.

\paragraph{Scalability, Customization, and Future Improvements}
Although we have designed our environment to be easily scalable and customizable, the attributes of objects and the interaction mechanisms between objects and agents currently rely on user-defined configurations. This customization process is subject to user preferences and research objectives. In the future, we plan to follow the approach outlined in~\cite{srivastava2021behavior} by leveraging WordNet~\cite{wordnet} to automatically generate object attributes and interaction methods. This will enable the rapid expansion of the environment’s object library while allowing users to further tailor interactions through direct modifications in the Python code.  

Our current environment adopts a text-based game engine where all physical and visual interactions are abstracted. Moving forward, we aim to extend our framework to support 3D assets and physical simulation engines, enhancing the realism of agent-environment interactions. This expansion will allow researchers to flexibly utilize our environment in both abstract and more concrete settings, depending on their experimental needs.

\paragraph{Testing of Reasoning Models}
Due to time and computational constraints, we did not test reasoning models such as GPT-O1 or Deepseek R1~\cite{deepseekai2025deepseekr1} in our experiments. Consequently, some unsolved cases in our study may be successfully addressed by more advanced reasoning models, potentially leading to agent behaviors that better resemble human decision-making. 




\bibliography{custom}

\appendix

\newpage

\section{Task Coordination Details}
\label{sec:appendix_a}

In Table~\ref{table:task_description}, we list the specific task descriptions used in our collaborative task-solving experiment. During the experiment, all five tasks were simultaneously input into the memory of all agents. Due to agent heterogeneity, certain task components require specific agents to complete or can be completed more efficiently by particular agents. Therefore, agents must determine which part of the task they should be responsible for and then take action simultaneously.

Hierarchically, the highest-level task is preparing the company event, under which the five tasks listed in \ref{fig:computer_class} serve as sub-tasks. Each task description further contains low-level tasks, such as the task Prepare Enough Tables and Chairs in the Event Area, which includes sub-tasks like Move 2 tables and 2 chairs to the open area 1 from other locations. These sub-tasks, in turn, may require multiple steps to complete.

This multi-level hierarchical task structure evaluates not only the agents’ planning abilities but also their collaboration and communication skills.

\begin{table*}[h!]
\centering
\caption{Task Descriptions}
\begin{tabularx}{\textwidth}{|c|X|X|}
\hline
\textbf{Task Number} & \textbf{Task Name} & \textbf{Description} \\
\hline
1 & Prepare Enough Tables and Chairs in the Event Area & Move 2 tables and 2 chairs to the open area 1 from other locations. \\
\hline
2 & Prepare Clean Utensils for the Event & Transport 4 clean plates, 4 clean knives, and 4 clean forks from other locations to the open area 1 and place them on the tables. \\
\hline
3 & Check and Repair Broken Computer, Projector, and Microphone & Move a podium to the open area 1 if there is no podium there. Bring one computer, one projector, and one microphone to the open area 1 and place them on the podium. Ensure they are in working condition. \\
\hline
4 & Book a Meeting Room for the Event & Use the touch screen in the open area 1 or a computer to remotely reserve the open area 1. Ensure the meeting room is booked with the following details: Event Name: Lunch and Listen; Start time: 2024-09-02T12:00:00; End time: 2024-09-02T13:00:00. \\
\hline
5 & Prepare Coffee, Tea, and Lunch for the Event & Take clean cups and use them to prepare 3 cups of coffee and 3 cups of tea and place them on the tables in the open area 1. Bring 2 meals to the open area 1 and place them on the tables in the open area 1. Ensure that meals are heated. \\
\hline
\end{tabularx}
\label{table:task_description}
\end{table*}
\section{Architect Survey on Multi–level Heterogeneity}
\label{sec:appendix_d}

\begin{figure*}[h]
  \centering
  \includegraphics[width=0.78\linewidth]{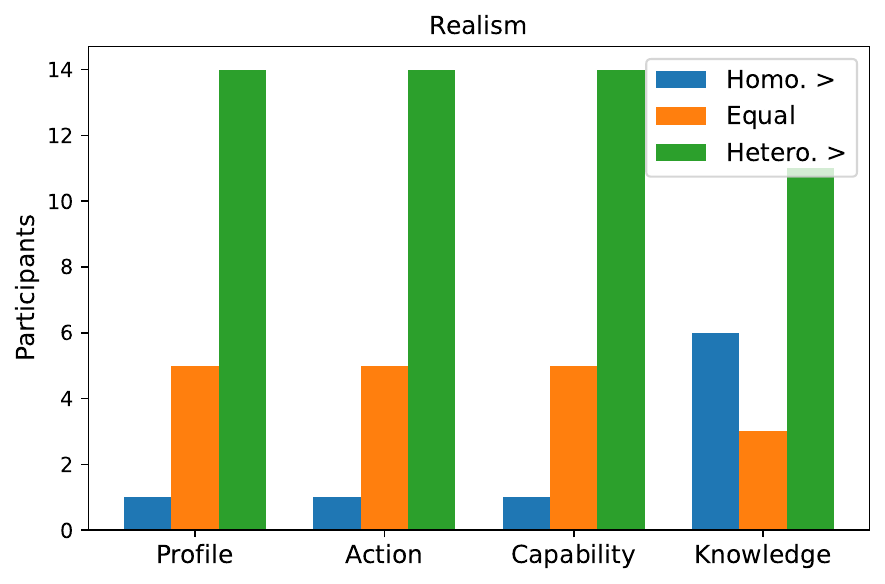}
  \caption{Realism comparison (\textbf{Homo.}=homogeneity, \textbf{Hetero.}=heterogeneity). A majority consider heterogeneity more realistic than homogeneity at every level.}
  \label{fig:realism}
\end{figure*}

\begin{figure*}[h]
  \centering
  \includegraphics[width=0.78\linewidth]{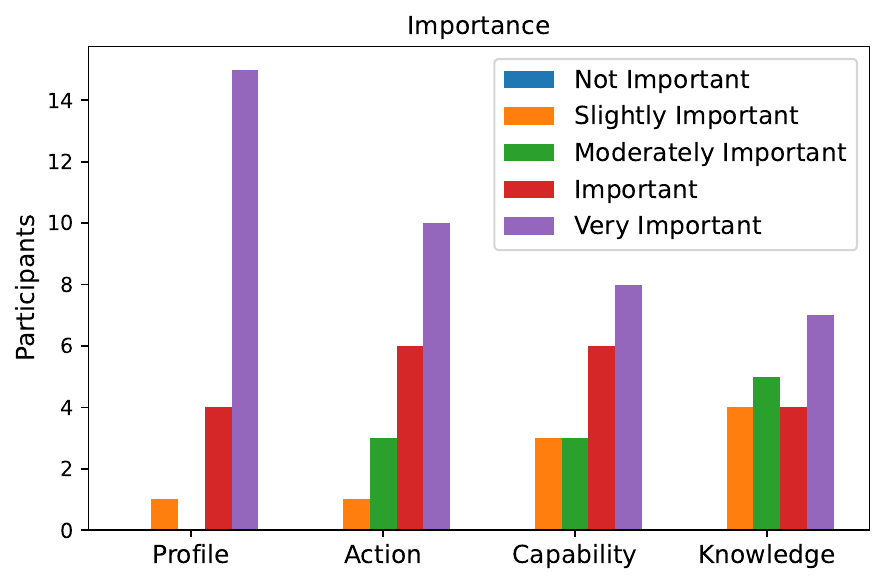}
  \caption{Importance ratings for understanding real space use (5-point scale). Most participants rate heterogeneity as “important” or “very important” at each level.}
  \label{fig:importance}
\end{figure*}

\begin{figure*}[h]
  \centering
  \includegraphics[width=0.78\linewidth]{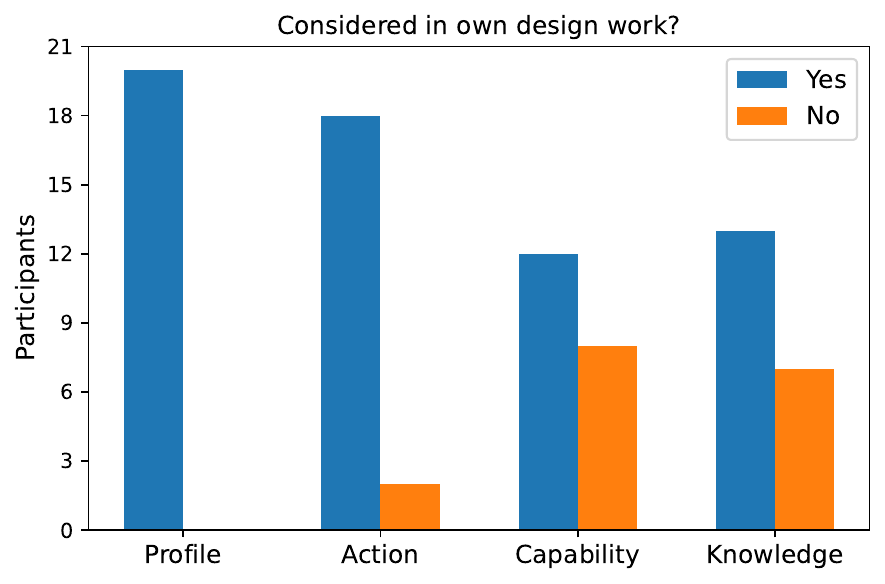}
  \caption{Whether architects typically consider each heterogeneity type in their
           own work.}
  \label{fig:consider}
\end{figure*}

\begin{figure*}[h]
  \centering
  \includegraphics[width=1\linewidth]{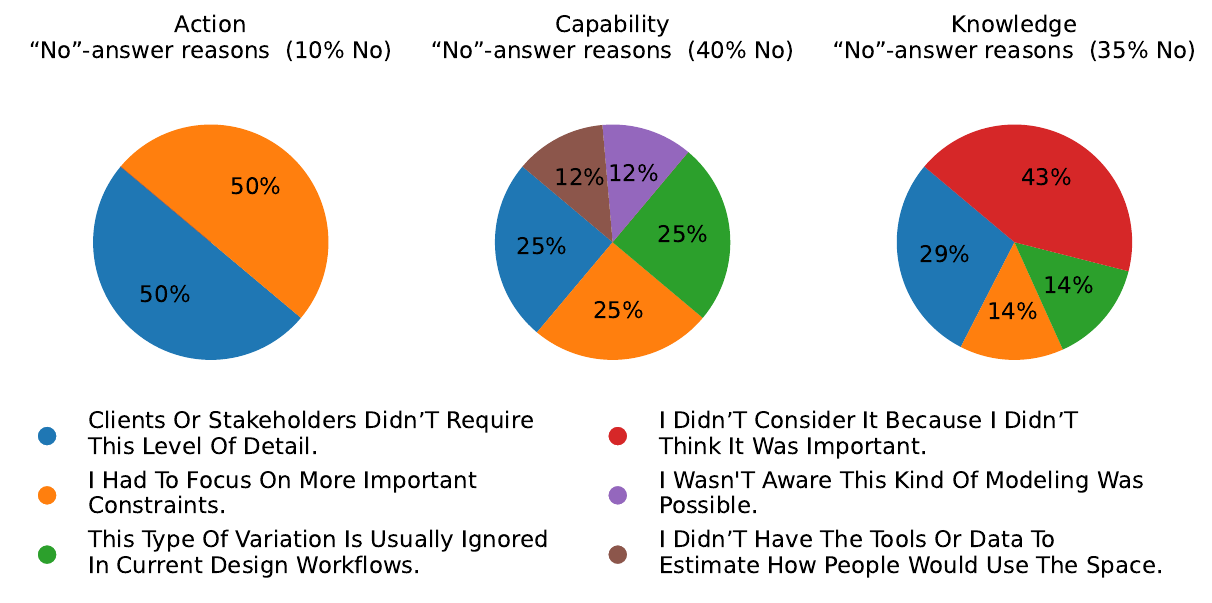}
  \caption{Reasons given by those who answered “No” in
           Fig.~\ref{fig:consider}.}
  \label{fig:reasons}
\end{figure*}
Table~\ref{tab:heterogeneity_comparison} compares the modeling of fine-grained heterogeneity types across four heterogeneous multi-agent environments. Our approach is the only one that comprehensively supports all four types of heterogeneity: profile, action, capability, and knowledge. To validate the necessity of modeling these levels of heterogeneity, we conducted a survey with 20 architects, asking for their views on the realism, importance, and consideration of heterogeneity modeling.

\begingroup
\setlength{\tabcolsep}{3.0pt}  
\renewcommand{\arraystretch}{1.0}  
\begin{table*}[ht]
\centering
\footnotesize  
\begin{tabular}{lcccc}
\toprule
\textbf{Environment} & \textbf{Profile} & \textbf{Action} & \textbf{Capability} & \textbf{Knowledge} \\
\midrule
D2A~\cite{wang2024simulating}                 & \yes & \no  & \no  & -- \\
Humanoid Agents~\cite{wang-etal-2023-humanoid}      & \yes & \no  & \no  & -- \\
Smallville~\cite{Park2023GenerativeAgents}    & \yes & \no  & \no  & \yes \\
Ours                 & \yes & \yes & \yes & \yes \\
\bottomrule
\end{tabular}
\caption{Comparison of fine-grained heterogeneity types across heterogeneous multi-agent environments. “Yes” indicates support for a type, “--” means not mentioned.}
\label{tab:heterogeneity_comparison}

\end{table*}
\endgroup

\begingroup
\setlength{\tabcolsep}{3pt}
\renewcommand{\arraystretch}{1}

\begin{table*}[th]
\centering
\footnotesize

\makebox[\textwidth][c]{%
  \begin{tabular}{llc}
  \toprule
  \textbf{ID} & \textbf{Question} & \textbf{Avg} \\
  \midrule
  \multicolumn{3}{l}{\textbf{Relevance of Design Problems}} \\[1pt]
  q1 & Resource allocation is a critical factor in office space planning. & 3.8 \\
  q2 & Spatial layout significantly affects occupant behavior and well-being. & 4.7 \\
  q3 & I consider human needs when designing building layouts. & 3.8 \\
  q4 & Evaluating waiting time or competition for shared resources is useful for improving space efficiency. & 4.0 \\
  \midrule
  \multicolumn{3}{l}{\textbf{Usefulness of \textsc{IndoorWorld}}} \\[1pt]
  q5  & The simulation of resource competition provides actionable insights for real-world resource placement. & 3.1 \\
  q6  & The spatial layout experiment helps reveal how different designs impact occupant behavior and efficiency. & 4.0 \\
  q7  & I can imagine using such simulations to evaluate and refine my own design decisions. & 3.9 \\
  q8  & The observed agent behavior reflects realistic office usage patterns. & 3.0 \\
  q9  & I would be interested in trying out this simulation tool with my own design layouts. & 4.1 \\
  q10 & I believe simulation tools like this can bridge the gap between architectural design and human behavioral modeling. & 4.2 \\
  \bottomrule
  \end{tabular}
} 

\caption{Average Likert ratings (1 = Strongly Disagree, 2 = Disagree, 3 = Neutral, 4 = Agree, 5 = Strongly Agree) provided by nine practicing architects, summarizing perceived relevance of key design problems and the usefulness of the proposed \textsc{IndoorWorld} simulation tool.}

\label{tab:design_relevance_usefulness}
\vspace{-4mm}
\end{table*}

\endgroup

\paragraph{Participants and Protocol}
We recruited participants through the \url{UserTesting.com} panel, targeting testers with professional architectural backgrounds, and compensated every tester at a rate above the statutory minimum wage applicable in their place of residence. Twenty practicing architects with at least 3 years’ experience completed an on-line questionnaire. For each of the four heterogeneity levels introduced in Section \ref{subsec:agent_arch}, respondents (i) compared two design options, \emph{A}: fully homogeneous occupants, and \emph{B}: heterogeneous occupants, and indicated which looked \emph{more realistic}; (ii) evaluated how \emph{important} the heterogeneity was for understanding real space use (5-point Likert scale); (iii) reported whether they normally consider that difference in their own design work, and, if not, \emph{why}.

\paragraph{Results Overview}
Figure \ref{fig:realism} shows that heterogeneity was judged more realistic by
55–70 \% of architects depending on the level, with the highest endorsement for
\emph{profile}, \emph{action} and \emph{capability} heterogeneity.  
Importance ratings (Fig.~\ref{fig:importance}) follow a similar trend:
55–95 \% of respondents marked the heterogeneity as “important” or “very important”. Figure~\ref{fig:consider} shows that most architects already incorporate heterogeneity consideration in their own workflow:
all 20 architects consider profile-level differences, and large majorities do so for
action (90\,\%), capability (60\,\%), and knowledge (65\,\%) heterogeneity.
This strong uptake suggests that heterogeneous agent modeling is both familiar
and valued in practice, particularly at the profile and action levels. However, 
some architects still do not consider these differences in their design 
work, citing various reasons. Fig.~\ref{fig:reasons} presents the further analysis of these reasons. For action-level heterogeneity, the two architects who do not consider it pointed to external factors: one mentioned that \emph{clients or stakeholders 
did not require this level of detail}, while the other cited \emph{having to 
focus on more important constraints}. At the capability level, the reasons were 
more diverse. While 25\,\% of those who opted out mentioned stakeholder 
requirements, another 25\,\% indicated that \emph{this type of variation is 
usually ignored in current design workflows}. An additional 25\,\% referenced 
focusing on more pressing design aspects, while the remaining 25\,\% were 
divided between \emph{not being aware this kind of modeling was possible} and 
\emph{lacking the tools or data to support it}. At the knowledge level, the 
primary reason for non-consideration was a perceived lack of importance. Other 
respondents mentioned stakeholder scope (29\,\%), and a smaller portion cited 
workflow limitations or practical constraints.

\paragraph{Implications}
This architect survey indicates that our multi-level heterogeneity yields more realistic simulations than existing environments and aligns more closely with the factors architects consider during design. This feature enhances the potential of \indoorworld as a supportive tool in architectural practice.

\section{Architect Survey on Design Problem Relevance and Tool Usefulness}
\label{appendix:survey_2}

\paragraph{Survey Design and Participants}  
We invited nine practicing architects (3–15 years of experience) to rate ten statements on a five-point Likert scale (1 = Strongly Disagree, 2 = Disagree, 3 = Neutral, 4 = Agree, 5 = Strongly Agree).  
The questionnaire comprised two blocks:

\begin{enumerate}
  \item \textbf{Relevance of Design Problems} (q1–q4): This section assessed the perceived importance of key design issues, including resource allocation, spatial layout, human-need considerations, and waiting-time analysis, in everyday architectural practice.
  \item \textbf{Usefulness of \textsc{IndoorWorld}} (q5–q10): This section evaluated whether the proposed simulation tool supports architects in understanding and addressing these design problems, as well as improving their design decisions.
\end{enumerate}

Questions and average scores are summarized in Table~\ref{tab:design_relevance_usefulness}.

\paragraph{Key Findings}  
The survey results indicate strong recognition of the relevance of the design problems addressed. All four items in the first block received high average ratings (M = 3.8–4.7). Notably, the importance of spatial layout (q2, M = 4.7) was overwhelmingly endorsed, reflecting its critical role in influencing occupant behavior and well-being. Similarly, the significance of analyzing waiting times or competition for shared resources (q4, M = 4.0) was highlighted, affirming the relevance of this problem to practical design.

For tool usefulness, respondents generally expressed positive views regarding the benefits of \textsc{IndoorWorld}. The simulation was appreciated for its ability to reveal layout impacts (q6, M = 4.0), support design refinement processes (q7, M = 3.9), and spark interest for direct use in design projects (q9, M = 4.1). Its potential to bridge the gap between architectural design and user behavior was also well-received (q10, M = 4.2). These responses suggest that architects see \textsc{IndoorWorld} as a promising tool for enhancing their design workflows, offering insights that can inform spatial planning and resource management.

However, two items received only neutral-to-slightly-positive ratings: actionable insights from resource-competition simulation (q5, M = 3.1) and the realism of observed agent behavior (q8, M = 3.0). These lower scores reflect a current gap between the simulation's fidelity and user expectations. Respondents generally acknowledged the concept of resource competition but found "hydration competition" less realistic compared to competition for meeting spaces or other workspace resources, which were viewed as more contextually relevant. One architect expressed concerns that the current game-like interface of \textsc{IndoorWorld}, where agents continuously move around and interact with each other, might deter some architects. They suggested that the simulation could be executed in the background, with results presented as concise, actionable recommendations for spatial planning, minimizing unnecessary visual complexity. Despite these reservations, respondents recognized that the simulation provides useful insights and proposed several enhancements. These included introducing more diverse agent profiles with varying needs and preferences to better assess universal design principles and accessibility, as well as enabling the visualization of agent movement flows to enhance spatial analysis.

\paragraph{Implications}  
Overall, the survey confirms that our modeling choices address key design challenges that architects recognize as important. The high scores for spatial layout and resource competition demonstrate that the simulated scenarios align well with real-world concerns. The positive feedback on tool usefulness suggests that \textsc{IndoorWorld} is perceived as a valuable support tool for architectural design, providing insights that can inform spatial planning, resource management, and design optimization.

Moreover, the relatively lower scores on agent behavior realism and the actionability of insights indicate areas for improvement. These results highlight a common challenge for many simulation tools: accurately capturing the complexity of human behavior while ensuring that the generated insights are directly applicable to design decisions. Currently, LLM-based agents are at an early stage of development, with simplified behaviors and limited contextual awareness. We anticipate that ongoing advancements in LLM capabilities will enhance agent realism, enabling more nuanced interactions and generating insights that align more closely with real-world user expectations. 



\section{Action-space comparison}
\label{appendix_f}

\begingroup
\setlength{\tabcolsep}{1.0pt}  
\renewcommand{\arraystretch}{1}
\begin{table}
\centering
\footnotesize  
\begin{tabular}{l|c}
\hline
 & \makecell{\#Actions} \\
\hline
\makecell[l]{Virtual-Home~\cite{puig2018virtualhome}} & 18 \\
\makecell[l]{ALFWorld~\cite{ALFWorld20}} & 9 \\
\makecell[l]{TDW-MAT~\cite{zhang2024building}} & 7 \\
\makecell[l]{C-WAH~\cite{zhang2024building}} & 8 \\
\makecell[l]{ALFRED~\cite{ALFRED20}} & 13 \\
\makecell[l]{TDW~\cite{gan2021threedworld}} & 12 \\
\makecell[l]{AdaSociety~\cite{huang2025adasociety}} & 12 \\
\hline
\makecell[l]{\textbf{Ours}} & \textbf{38} \\
\hline
\end{tabular}
\caption{Comparison of the number of actions supported by different environments.}
\label{tab:number_of_actions}
\vspace{-5mm}
\end{table}
\endgroup

Table~\ref{tab:number_of_actions} summarizes the size of the action space
supported by several widely used task-oriented simulation environments.
Most benchmarks provide fewer than 20 primitive actions:
VirtualHome offers 18, ALFWorld 9, and the TDW variants between 7 and 12.
Our environment markedly expands this spectrum to \textbf{38 distinct actions},
more than twice that of any prior system listed.
The richer verb set enables agents to engage in a broader range of household
manipulations (e.g., \texttt{brewCoffee}, \texttt{repairDevice},
\texttt{heatFood}), which in turn supports more realistic task decompositions,
greater behavioral diversity, and nuanced capability differences.

\section{Environment Initialization via JSON}
\label{sec:appendix_b}

Our environment is defined using a JSON configuration file, which specifies key components such as agents, locations, inter-location connections, receptacles within locations, and objects. This structured definition allows for flexible customization of the simulation environment.

Our code automatically initializes the environment based on the JSON file, creating corresponding instances of \texttt{Location}, \texttt{Receptacle}, \texttt{Object}, and \texttt{Agent} classes. Each element in the JSON file is mapped to its respective class, ensuring that all objects and agents are correctly instantiated with their predefined attributes and relationships.

An example of this JSON configuration is shown in Figure~\ref{fig:json_config}, where locations, objects, receptacles, and agents are defined. This JSON-based approach enables researchers to easily modify and extend the environment without changing the core simulation code, making it a highly adaptable framework for various research scenarios.

\section{Extending Agent-Object Interactions}
\label{sec:appendix_c}

\begin{figure*}[h]
\centering
\begin{lstlisting}[language=json, caption={}, label={lst:json}]
{
  "locations": ["kitchen", "meeting_room1"],
  "location_distances": {
    "kitchen": {"meeting_room1": 1},
    "meeting_room1": {"kitchen": 1}  
  },
  "receptacles": [
    {"name": "Sinkbasin1", "location": "kitchen", "rtype": "Sinkbasin", 
     "weight_kg": 15, "state": {"fixed": true, "closable": false, "is_open": true, 
     "is_clean": true, "temperature": 20, "is_working": true}},
    {"name": "Cabinet1", "location": "kitchen", "rtype": "Cabinet", 
     "weight_kg": 30, "state": {"fixed": true, "closable": true, "is_open": false, 
     "is_clean": true, "temperature": 20, "is_working": true}}
  ],
  "objects": [
    {"name": "touchscreen_1", "otype": "TouchScreen", "location": "meeting_room1", 
     "weight_kg": 3, "state": {"is_turned_on": true, "is_working": true, "is_clean": true, 
     "temperature": 20}},
    {"name": "cup_1", "otype": "Cup", "location": "kitchen", "receptacle": "Countertop1", 
     "weight_kg": 0.3, "state": {"is_clean": false, "temperature": 20}}
  ],
  "agents": [
    {
      "name": "ryan", 
      "gender": "male", 
      "role": "receptionist",
      "location": "meeting_room1", 
        "fullness": 100, 
        "hydration": 100, 
        "energy": 100, 
        "social_fulfillment": 100, 
      "strength_kg": 65, 
      "internal_profile": "Ryan is a professional and welcoming receptionist. 
        Known for his friendly personality and exceptional communication skills....",
      "appearance": "Ryan is a receptionist who is tall and well-built, with ..."
    },
    {
      "name": "irene", 
      "gender": "female", 
      "role": "IT_admin",
      "location": "kitchen", 
        "fullness": 100, 
        "hydration": 100, 
        "energy": 100, 
        "social_fulfillment": 100, 
      "strength_kg": 30,
      "internal_profile": "Irene is an organized and skilled IT administrator. 
        She is quick to troubleshoot and efficiently repair a wide range of ...",
      "appearance": "Irene has a petite, tidy appearance with a focused expression. 
        She is usually dressed casually but professionally, with an attentive ..."
    }
  ]
}
\end{lstlisting}
\caption{Excerpt from the JSON configuration file defining locations, agents, objects, and receptacles in the environment.}
\label{fig:json_config}
\end{figure*}
\begin{figure*}[h]
\centering
\begin{lstlisting}[style=python, caption={}, label={lst:computer}]
class Computer(BaseObject):
    def __init__(self, name, otype='Computer', location, environment, weight_kg, 
                 carryable=False, requires_receptacle=True, 
                 state={"is_clean": True, 'temperature': 20}):
        super().__init__(name, otype, location, environment, weight_kg, 
                         carryable, requires_receptacle, state=state)
    def power_on(self):
        if self.state['is_turned_on']:
            return f"{self.name} is already turned on.", False
        self.state['is_turned_on'] = True
        return f"{self.name} is now turned on.", True

    .....
        
    def get_admissible_actions(self, agent):
        actions = super().get_admissible_actions(agent)

        # Add repair action
        if f'repair_{self.otype.lower()}' in agent.skills:
            actions.append(f"repair_{self.otype.lower()} {self.name}")

        # Check if the computer has an 'owner' attribute
        if self.location == agent.location:
            if not self.state['is_turned_on']:
                actions.append(f"turn_on {self.name}")
            else:
                actions.append(f"turn_off {self.name}")

        return actions
\end{lstlisting}
\caption{Illustration of the \texttt{Computer} class, defining interaction methods like \texttt{power\_on()} and specifying admissible agent actions.}
\label{fig:computer_class}
\end{figure*}

\begin{figure*}[h]
\centering
\begin{lstlisting}[style=python, caption={}, label={lst:agent}]
class Agent:
    def __init__(self, name, role, location, skills=None):
        self.name = name
        self.role = role
        self.location = location
        self.skills = skills if skills else {}

    def act(self, command):
        """Execute an action if it's admissible."""
        command_mapping = {
            "turn_on": (self.turn_on,1), # 1 means one argument
            ...

        }

        tokens = command.split()
        if len(tokens) >= 1:
            action = tokens[0]

            if action in command_mapping:
                action_func, arg_count = command_mapping[action]
                if len(tokens[1:]) >= arg_count:
                    args = tokens[1:1 + arg_count]
                    return action_func(*args)
                else:
                    return f"{self.name} received an incorrect number of arguments for action {action}.", False
            else:
                return f"{self.name} cannot perform action {action}.", False

        return f"{self.name} cannot parse command {command}.", False

    def turn_on(self, electronic_device_name):

    
        # Search for the electronic device in the current location's objects
        target_device = next((obj for obj in self.location.objects if obj.name == electronic_device_name), None)
        
        # If the device is found, attempt to turn it on
        if target_device:
            if hasattr(target_device, "turn_on"):
                return target_device.turn_on()
            else:
                return f"{target_device.name} cannot be turned on.", False
        else:
            return f"{self.name} cannot find {electronic_device_name}.", False
            
    def get_admissible_actions(self):
        admissible_actions = []
        ...
        if self.location.objects:
            for obj in self.location.objects:
                if obj.receptacle == None or obj.receptacle.state['is_open']:
                    admissible_actions.extend(obj.get_admissible_actions(self))
        ...
        return admissible_actions
\end{lstlisting}
\caption{Illustration of the \texttt{Agent} class, where \texttt{act()} maps commands to interaction methods like \texttt{turn\_on()}.}
\label{fig:agent_act}
\end{figure*}

\begin{figure*}[ht]
\centering
\begin{lstlisting}[style=python, caption={}, label={lst:itadmin_repair}]
class ITAdmin(Agent):
    def __init__(self, name, gender, location, environment, **kwargs):
        super().__init__(name, gender, location, environment, **kwargs)
        # ITAdmin specific skills for repairing devices
        self.skills.update({
            "repair_computer": 1,
            ...
        })

    def repair_electronic_device(self, obj_name):
        # Find the object with the specified name in the current location
        target_device = next((obj for obj in self.location.objects if obj.name == obj_name), None)

        if target_device:
            # Check if the object is a repairable electronic device
            electronic_devices = ["Microphone", "Projector", "Computer", "CoffeeMachine", "WaterDispenser", "Microwave"]
            if target_device.otype in electronic_devices:
                # Check if the device is broken
                if "is_working" in target_device.state and not target_device.state["is_working"]:
                    target_device.state["is_working"] = True
                    return f"{self.name} repaired the {target_device.name}.", True
                else:
                    return f"{target_device.name} is already in working condition.", False
            else:
                return f"{target_device.name} is not a repairable electronic device.", False
        else:
            return f"{self.name} cannot find {obj_name} in the current location.", False

    def act(self, command):
        """Execute an ITAdmin-specific action or fall back to the Agent's act method."""
        command_mapping = {
            "repair_computer": (self.repair_electronic_device, 1),
            ...
        }

        tokens = command.split()
        if len(tokens) >= 1 and tokens[0] in command_mapping:
            action_func, arg_count = command_mapping[tokens[0]]
            if len(tokens[1:]) >= arg_count:
                return action_func(*tokens[1:1 + arg_count])
            return f"{self.name} received an incorrect number of arguments.", False

        return super().act(command)
\end{lstlisting}
\caption{Illustration of the \texttt{ITAdmin} class, where \texttt{repair\_computer()} enables role-specific actions via command mapping.}
\label{fig:itadmin_repair}
\end{figure*}

Figures~\ref{fig:computer_class}, \ref{fig:agent_act}, and \ref{fig:itadmin_repair}
provide three example classes that illustrate how researchers can extend the
environment by adding new object types and by defining interactions between
agents and those objects.

For example, if a researcher wants to introduce a new class, such as \texttt{Computer}, they first need to specify what actions agents can perform on it, such as \texttt{turn\_on}. This is done by defining interaction methods like \texttt{power\_on()} within the \texttt{Computer} class. The new class should inherit from \texttt{BaseObject} and implement \texttt{get\_admissible\_actions()}, which determines when and how agents can interact with the object. For instance, if the computer is in an accessible state, it can return a command like \texttt{turn\_on \{self.name\}}.

On the agent side, new interaction methods must be defined accordingly. For instance, the \texttt{Agent} class should implement a \texttt{turn\_on()} method that handles the interaction logic for powering on a \texttt{Computer}. Additionally, in the \texttt{act()} function, a new command mapping should be added, linking the \texttt{turn\_on} command to the \texttt{turn\_on()} method. This way, when the agent's \texttt{get\_admissible\_actions()} method runs, it will include the new \texttt{turn\_on} action if the object is in the same location as the agent and is not placed inside a closed receptacle. The agent can then decide whether to execute this action, effectively enabling interaction with the newly introduced \texttt{Computer} object.

The \texttt{ITAdmin} class further demonstrates how to define role-specific actions. This is done by first adding \texttt{"repair\_computer"} to the \texttt{skills} dictionary. Then, a new method, such as \texttt{repair\_electronic\_device()}, is defined, and the \texttt{act()} function maps the \texttt{"repair\_computer"} command to this method. This allows the \texttt{ITAdmin} role to perform repair actions that are unavailable to other agents.

\textbf{Note:} These examples have been simplified for clarity and may not exactly match the original source code. Researchers should refer to the actual implementation details to fully integrate new objects and interactions into the environment.

\end{document}